\documentclass[12pt]{article}

\usepackage{setspace} 
\usepackage{placeins}
\usepackage{comment}

\usepackage[margin=1in]{geometry}
\usepackage{amsmath,amssymb,amsthm,mathtools}
\usepackage{enumitem}
\usepackage{hyperref}
\usepackage{tikz}
\usepackage{pgfplots}
\pgfplotsset{compat=1.18}
\usetikzlibrary{arrows.meta}
\usepgfplotslibrary{groupplots}

\hypersetup{colorlinks=true,linkcolor=blue,citecolor=blue,urlcolor=blue}

\usepackage{natbib,hyperref}
\newtheorem{assumption}{Assumption}
\newtheorem{lemma}{Lemma}
\newtheorem{theorem}{Theorem}

\newtheorem{corollary}{Corollary}

\newcommand{\E}{\mathbb{E}}

\newcommand{\pto}{\stackrel{p}{\to}}

\newcommand{\eps}{\varepsilon}

\title{A Tail-Geometric Approach to Kotlarski’s Lemma with Order Statistics\thanks{
Contact Information: Department of Economics, University of Toronto, Max Gluskin House, 150 St. George St, Toronto, ON M5S 3G7, Canada (email: yao.luo@utoronto.ca). We thank Bingyao Liu for excellent research assistance. All remaining errors are our own.
}}

\author{Yao Luo}

\date{November 2026}

\begin{document}
\maketitle

\begin{abstract}
We develop a tail-geometric approach to nonparametric identification in repeated-measurement models based on large within-unit gaps. Differencing eliminates the common latent component, while conditioning on a large gap pushes the idiosyncratic errors into opposite tails. Under a regular thinner-than-Gaussian tail condition, an idiosyncratic component concentrates around a deterministic, gap-dependent location. Consequently, the corresponding observed measurement, after recentering, reveals the distribution of the common latent component up to location. The idiosyncratic distribution can then be recovered through deconvolution. This principle applies both to the classical two-measurement Kotlarski model and to settings where only order statistics are observed. Applications include auctions and wage-offer models, whether all offers are observed or only the top-ranked ones are available.
\end{abstract}


\vspace{.5 cm}


\newpage 

\section{Introduction}

Kotlarski's lemma (\citet{Kotlarski1967}) provides a fundamental identification result for repeated-measurement models. When two measurements share a common latent component but contain independent idiosyncratic errors, their joint distribution identifies the distributions of the common and idiosyncratic components up to location. Since its introduction to econometrics by \citet{LiVuong1998}, this result has become an important tool for separating latent heterogeneity from idiosyncratic variation. Applications include empirical auctions (\citet{LiPerrigneVuong2000}; \citet{Krasnokutskaya2011}; \citet{GrundlZhu2019}), earnings dynamics (\citet{BonhommeRobin2010}; \citet{HuMoffittSasaki2019}), and models of skill formation and other latent heterogeneity (\citet{cunha2010estimating}). See \citet{Schennach2016} for a review.

In many economic settings, however, researchers do not observe the repeated measurements themselves, but only selected order statistics. This occurs naturally in auctions, where ascending-auction formats and data-recording practices may reveal only some of the highest bids, and in wage-offer data, where surveys may record only an individual's best offers; see, e.g., \citet{AtheyHaile2002}, \citet{hernandez2020estimation}, and \citet{Guo2025}. Because order statistics are dependent even when the underlying idiosyncratic errors are independent, the independence structure underlying the classical Kotlarski argument is no longer available. This raises a general question: can the latent and idiosyncratic distributions still be recovered when the researcher observes only order statistics of the measurements?

A natural first step is to difference the observed order statistics. Their spacing eliminates the common latent component exactly. This observation alone, however, does not solve the identification problem. As shown by \citet{rossberg1972characterization}, distinct parent distributions can generate the same distribution of the spacing between two i.i.d.\ observations. Thus, although differencing removes the common component, the spacing distribution alone generally discards too much information to recover the idiosyncratic distribution. Identification requires exploiting additional information contained in the joint distribution of the observed order statistics.

This paper develops a new identification argument based on extreme spacings. The central idea is that, conditional on an extreme spacing, support or tail geometry sharply restricts where the underlying idiosyncratic order statistics can be located. With finite support, a spacing approaching its maximal feasible value pushes the two order statistics toward opposite support boundaries. With full support, a large spacing instead pushes them into opposite tails, where their locations are determined by tail geometry. Under regular thinner-than-Gaussian tails, we show that the relevant idiosyncratic order statistic concentrates around a deterministic large-gap location. Subtracting this location from the corresponding observed order statistic therefore reveals the distribution of the common latent component. Once the latent distribution is identified, the distribution of the idiosyncratic order statistic follows by deconvolution, and the parent error distribution can then be recovered from the one-to-one relationship between an order-statistic distribution and its parent distribution.

More broadly, our approach provides a unified geometric
principle for identification from classical repeated
measurements and order statistics across different support
configurations. Its main new implication is that two order
statistics suffice for nonparametric identification even
when the idiosyncratic distribution has unbounded support
in both directions.\footnote{Unbounded support can arise naturally in auction applications even when values and bids themselves are positive. For example, bidder values are often modeled as the product of auction-level heterogeneity and bidder-specific heterogeneity. Taking logs converts this multiplicative structure into the additive form studied here. Even if the bidder-specific component is strictly positive, its logarithm can be arbitrarily negative when that component can be arbitrarily close to zero.} The approach also provides an
alternative geometric identification argument for the
classical two-measurement Kotlarski model.

The paper contributes to the literature on identification of auction models with unobserved heterogeneity using measurement-error and related approaches. One strand extracts information from limited auction records or a single observed order statistic. \citet{kong2022identification} establishes identification in English auctions when losing entrants are unobserved, \citet{guerre2019nonparametric} exploit density discontinuities in winning bids generated by unobserved competition, and \citet{bugni2023inference} develop inference for auctions with many bidders using transaction prices alone. 

Another strand exploits multiple measurements or order statistics to separate common and idiosyncratic heterogeneity, including \citet{LiPerrigneVuong2000} and \citet{hu2013identification}. In particular, \citet{LuoXiao2023} exploit the
multiplicatively separable structure of the joint
density of consecutive order statistics to identify
auction models with discrete, potentially nonseparable
unobserved heterogeneity, using additional information
from an instrument.\footnote{See also \citet{luo2020unobserved} and \citet{luo2024order}.} In contrast, we consider an additive separable model and exploit large-gap
concentration. Our approach does not require an instrument and
extends to nonconsecutive order statistics, at the
cost of imposing restrictions on tail behavior.

Most closely related is \citet{ChoLuoXiao2024}, who
establish identification from two order statistics
under a finite support boundary. Their approach
exploits characteristic-function restrictions involving
spacings and cross-sums to identify the idiosyncratic
distribution, followed by deconvolution to recover
the common latent distribution. In contrast, our
approach uses large-gap conditioning to identify
the common latent distribution first, which then
allows the idiosyncratic distribution to be recovered.
Importantly, this alternative argument establishes
identification even when both tails of the
idiosyncratic distribution are unbounded.

Our approach also complements extensions of Kotlarski's lemma. Such extensions typically retain the repeated-measurement structure while relaxing characteristic-function conditions or modifying the way latent components enter the measurements; see, for example, \citet{EvdokimovWhite2012}, \citet{LiZheng2020}, \citet{Lewbel2022}, and \citet{franguridi2026kotlarski}. Our approach instead replaces the characteristic-function factorization with an argument based on the conditional geometry of extreme spacings.

The role of extreme spacings in our proposed approach is primarily to establish identification. Estimation need not rely only on observations with very large gaps. Once the model is identified, the unknown distributions determine the joint distribution of the observed order statistics at all spacing values, so sieve-based likelihood or minimum-distance methods can in principle use the entire joint distribution. See, e.g., \citet{chen2007large}. For example, \citet{ChoLuoXiao2024} develop a simulated sieve estimator based on the joint
distribution of observed order statistics. Our extreme-spacing argument isolates a transparent source of identifying variation rather than prescribing that estimation use only tail observations.\footnote{We do not study estimation or inference in this paper. See, e.g., \citet{LiVuong1998}, \citet{menzel2013large}, \citet{kato2021robust}, \citet{kurisu2022uniform} and \citet{chen2025faster} for discussions of related settings.}

The remainder of the paper is organized as follows.
Section~2 introduces the model and develops the geometric
intuition behind the large-gap approach. Section~3
establishes the main concentration and identification
results under thinner-than-Gaussian tails, followed by
extensions to nonconsecutive order statistics and
one-sided bounded support. Section~4 concludes.

\section{Setup and Geometric Intuition}

Consider the repeated-measurement model
\[
X_j=\xi+\eta_j,\qquad j=1,\ldots,n,
\]
where $\xi$ is independent of the idiosyncratic components
$\eta_1,\ldots,\eta_n$, which are i.i.d.\ with distribution
function $F$ and density $f$. We impose the location normalization
\[
\E[\eta_j]=0.
\]
The density $f$ is continuous and strictly positive on the
interior of its support.

When at least two individual measurements are observed,
Kotlarski's lemma (\citet{Kotlarski1967}) identifies the
distributions of $\xi$ and $\eta_j$ nonparametrically under
standard regularity conditions. The classical argument
exploits the independence of the idiosyncratic components
through the joint characteristic function of the measurements.
By differentiating its logarithm and imposing a location
normalization, one recovers the characteristic function of
the common component. The idiosyncratic distribution then
follows by deconvolution.

In applications such as auctions and wage-offer models,
however, researchers may observe only selected order
statistics rather than individual measurements. Let
$X_{(j)}$ and $\eta_{(j)}$ denote the $j$-th smallest order
statistics of the measurements and idiosyncratic components,
respectively. Since the common component $\xi$ does not
affect their ranking,
\[
X_{(j)}=\xi+\eta_{(j)}.
\]
We ask whether the distributions of $\xi$ and $\eta_j$
remain identified when only a pair of order statistics,
\[
(X_{(r)},X_{(s)}),\qquad 1\leq r<s\leq n,
\]
is observed. We assume throughout that the sample size \(n\) and the ranks \(r,s\) of the observed order statistics are known.

For example, in an ascending private-value auction,
dropout prices reveal ordered bidder valuations,
but the highest valuation is typically unobserved.
When the two highest dropout prices are recorded,
the researcher observes $(X_{(n-2)},X_{(n-1)})$.
Our identification results apply to such partially
observed order-statistic data without requiring
the highest underlying valuation to be observed.

Unlike the original idiosyncratic components,
$\eta_{(r)}$ and $\eta_{(s)}$ are dependent, so the
characteristic-function factorization underlying Kotlarski's
lemma no longer applies. The closest result, \citet{ChoLuoXiao2024},
establishes identification from two order statistics
under a finite lower support boundary. Their argument
exploits characteristic-function restrictions involving
spacings and cross-sums, together with the finite
boundary, to recover the idiosyncratic distribution.
Our approach instead uses large-gap concentration,
which also applies when the idiosyncratic distribution
is unbounded in both directions.

\subsection{An Illustration in the Kotlarski Model}  \label{sec:example}

We first illustrate our idea in the classical two-measurement model,
\[
X_1=\xi+\eta_1,
\qquad
X_2=\xi+\eta_2.
\]

Consider first the case of compact error support, say $[\underline\eta,\bar\eta]$. Then if \(D =X_2 - X_1 = \eta_{2}-\eta_{1} \) approaches its maximal feasible value
\(\bar{\eta}-\underline{\eta}\),
\[
\eta_1 \to \underline{\eta},
\qquad
\eta_2 \to \bar{\eta},
\]
and hence
\[
X_2-\bar{\eta}\mid D=d
\Rightarrow \xi
\qquad
\text{as }
d\uparrow \bar{\eta}-\underline{\eta}.
\]
After subtracting the boundary point $\bar\eta$, the conditional distribution of $X_{2}$ reveals the distribution of $\xi$. This identifies the distribution of the common latent component up to location. The idiosyncratic error distribution can then be recovered, up to the opposite location shift, by deconvolving the marginal distribution of $X_{2}$. 

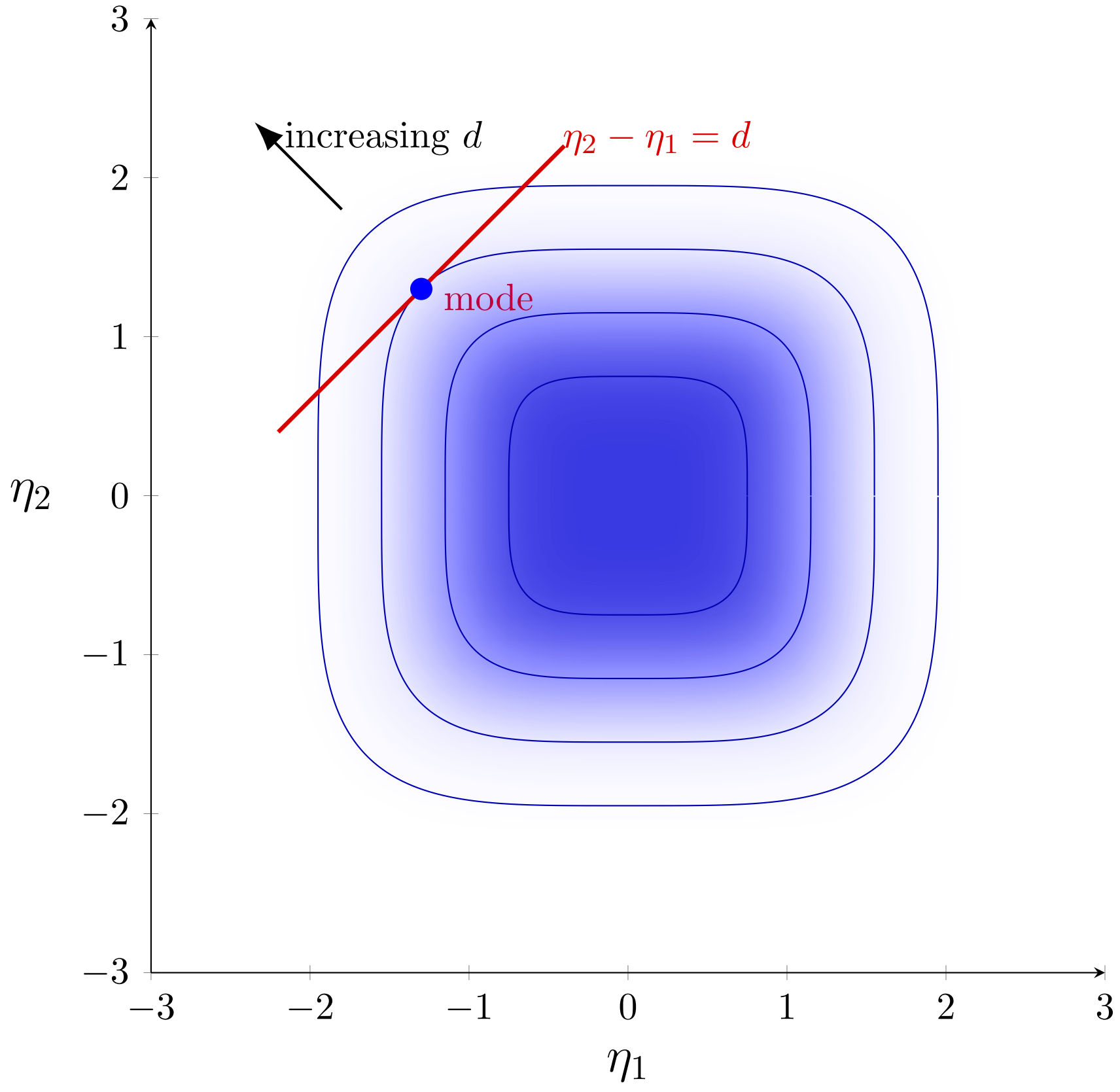
\begin{figure}[htbp]
\centering
\begin{tikzpicture}
\begin{axis}[
    width=0.88\textwidth,
    height=0.68\textwidth,
    xmin=-3, xmax=3,
    ymin=-3, ymax=3,
    axis equal image,
    xlabel={$\eta_1$},
    ylabel={$\eta_2$},
    xlabel style={font=\large},
    ylabel style={font=\large, rotate=-90},
    tick label style={font=\small},
    axis lines=left,
    enlargelimits=false,
    clip=true,
    view={0}{90},
    colormap={softblue}{
        color(0cm)=(white);
        color(0.55cm)=(blue!10);
        color(1.1cm)=(blue!30);
        color(1.8cm)=(blue!60);
        color(2.5cm)=(blue!85!black)
    },
    colorbar=false,
]

\pgfmathsetmacro{\sigmaeta}{1.00}
\pgfmathsetmacro{\dOne}{2.60}

\addplot3[
    surf,
    shader=interp,
    draw=none,
    opacity=0.78,
    domain=-3:3,
    y domain=-3:3,
    samples=41,
    samples y=41,
]
{exp(-(x^4+y^4)/(4*(\sigmaeta)^4))};

\foreach \rr in {0.75,1.15,1.55,1.95}{
    \addplot[
        blue!70!black,
        thin,
        domain=0:360,
        samples=240,
        smooth,
    ]
    (
        {\rr*sign(cos(x))*sqrt(abs(cos(x)))},
        {\rr*sign(sin(x))*sqrt(abs(sin(x)))}
    );
}

\addplot[
    red!85!black,
    very thick,
    domain=-2.2:-0.4,
    samples=2,
]
{x+\dOne};

\addplot[
    only marks,
    mark=*,
    mark size=3pt,
    blue,
]
coordinates {(-\dOne/2,\dOne/2)};

\node[red!85!black, font=\small, anchor=west]
    at (axis cs: -0.5,2.25)
    {$\eta_2-\eta_1=d$};

\node[purple, font=\small, anchor=west]
    at (axis cs:-1.25,1.25)
    {mode};

\draw[-{Latex[length=3mm]}, thick, black]
    (axis cs:-1.80,1.80) -- (axis cs:-2.35,2.35);

\node[font=\small, anchor=west]
    at (axis cs:-2.25,2.25)
    {increasing $d$};

\end{axis}
\end{tikzpicture}

\caption{Large-gap conditioning under thinner-than-Gaussian errors.}
\label{fig:kotlarski-large-gap-contour}

\vspace{0.4em}
\begin{minipage}{0.92\textwidth}
\footnotesize
\emph{Note.}
The heat map and contours show the joint density of
$(\eta_1,\eta_2)$ when
$f(\eta)\propto\exp\{-\eta^4/(4)\}$.
The red line is the conditioning event
$\eta_2-\eta_1=d$.
The blue point marks the mode of the joint density restricted to this line,
which is
$(-d/2,d/2)$.
As $d$ increases, the conditioning line moves toward the northwest.
The curvature of the conditional density around the constrained mode increases
with $d$, so that
$\eta_2-d/2\mid D=d\xrightarrow{p}0$.
\end{minipage}
\end{figure}

The same idea extends to full support. Instead of using a finite endpoint $\bar\eta$, we use the fact that under thinner-than-Gaussian tail curvature as defined below, conditioning on a large spacing makes $\eta_{2}$ stable after recentering at a deterministic large-gap center. Figure \ref{fig:kotlarski-large-gap-contour} illustrates the idea using a simple example where
\[
f(\eta) \propto
\exp\left\{
-\frac{\eta^4}{4}
\right\} .
\]
The gap \(X_2-X_1\) eliminates the common component. Conditioning on a value of the
gap restricts the joint distribution to a diagonal line, and the constrained
mode along this line determines the large-gap centering. We now show that the conditional distribution around this center concentrates as the gap grows.

Conditional on \(D=d\), the density of \(\eta_2\) is proportional to
\[
\exp\left\{
-\frac{(y-d)^4+y^4}{4}
\right\}.
\]
The detailed derivations are provided in Appendix~\ref{app:quartic}.

First, this conditional density has a unique mode,
\[
y_d=\frac{d}{2}.
\]
This is visible in Figure \ref{fig:kotlarski-large-gap-contour}: along the conditioning line, the point
\[
\left(-\frac{d}{2},\frac{d}{2}\right)
\]
lies in the darkest part of the heat map and therefore corresponds to the conditional mode.

Second, after recentering around the mode, the density of
\(\eta_2-\frac{d}{2}\) is proportional to
\[
\exp\left\{
-\frac{3d^2z^2}{4}
-\frac{z^4}{2}
\right\}.
\]
Figure \ref{fig:kotlarski-large-gap-contour} also illustrates this concentration: as the conditioning line moves outward with larger \(d\), the density falls off more sharply away from the constrained mode. Thus,
\[
\eta_2-\frac{d}{2}\mid D=d
\xrightarrow{p}0.
\]
Large-gap conditioning therefore asymptotically pins down \(\eta_2\) around its conditional mode, so that
\[
\xi \approx X_2-\frac{d}{2}
\]
for large \(d\). Thus, the gap-dependent mode plays the same role as the finite support boundary in the compact-support case, allowing the common latent distribution to be recovered up to location.

%
%

\pgfmathdeclarefunction{qgap}{2}{%
  \pgfmathparse{exp(-(#1)^4/2-3*(#2)^2*(#1)^2/4)}%
}

\pgfmathdeclarefunction{phi}{1}{%
  \pgfmathparse{exp(-(#1)^2/2)/sqrt(2*pi)}%
}


\pgfmathdeclarefunction{gapZ}{1}{%
  \begingroup
  \pgfmathsetmacro{\scale}{1/sqrt(1+1.5*(#1)^2)}%
  \def\acc{0}%
  \foreach \k in {0,...,24}{%
    \pgfmathparse{\acc +
      ifthenelse(\k==0 || \k==24,1,
        ifthenelse(mod(\k,2)==0,2,4)) *
      qgap(\scale*(-5+10*\k/24),#1)}%
    \global\let\acc\pgfmathresult
  }%
  \pgfmathparse{\acc*\scale*10/72}%
  \pgfmathsmuggle\pgfmathresult
  \endgroup
}

\pgfmathdeclarefunction{gapH}{2}{%
  \begingroup
  \pgfmathsetmacro{\scale}{1/sqrt(1+1.5*(#2)^2)}%
  \def\acc{0}%
  \foreach \k in {0,...,24}{%
    \pgfmathparse{\acc +
      ifthenelse(\k==0 || \k==24,1,
        ifthenelse(mod(\k,2)==0,2,4)) *
      phi((#1)-\scale*(-5+10*\k/24)) *
      qgap(\scale*(-5+10*\k/24),#2)}%
    \global\let\acc\pgfmathresult
  }%
  \pgfmathparse{\acc*\scale*10/72}%
  \pgfmathsmuggle\pgfmathresult
  \endgroup
}


\begin{figure}[htbp]
\centering

\pgfmathparse{gapZ(1)}\let\Zone\pgfmathresult
\pgfmathparse{gapZ(2)}\let\Ztwo\pgfmathresult
\pgfmathparse{gapZ(4)}\let\Zfour\pgfmathresult

\begin{tikzpicture}
\begin{groupplot}[
    group style={
        group size=2 by 1,
        horizontal sep=1.2cm
    },
    width=0.43\textwidth,
    height=0.36\textwidth,
    axis lines=left,
    grid=major,
    grid style={black!8},
    no markers,
    tick label style={font=\scriptsize},
    label style={font=\small},
    title style={font=\small},
    legend style={
        font=\scriptsize,
        draw=none,
        fill=none
    },
    samples=65,
]

\nextgroupplot[
    title={(a) Conditional error concentration},
    xlabel={$z=\eta_2-d/2$},
    ylabel={Conditional density},
    xmin=-1.6, xmax=1.6,
    ymin=0,
    domain=-1.6:1.6,
    legend pos=north east,
]

\addplot[blue!40!black,dashed,thick]
    {qgap(x,1)/\Zone};
\addlegendentry{$d=1$}

\addplot[blue!60!black,dashdotted,thick]
    {qgap(x,2)/\Ztwo};
\addlegendentry{$d=2$}

\addplot[blue!80!black,solid,very thick]
    {qgap(x,4)/\Zfour};
\addlegendentry{$d=4$}

\nextgroupplot[
    title={(b) Revealing the latent distribution},
    xlabel={$x=X_2-d/2$},
    ylabel={Conditional density},
    xmin=-3, xmax=3,
    ymin=0, ymax=0.45,
    domain=-3:3,
    legend pos=north east,
]

\addplot[black,densely dotted,very thick]
    {phi(x)};
\addlegendentry{$f_\xi$}

\addplot[blue!40!black,dashed,thick]
    {gapH(x,1)/\Zone};
\addlegendentry{$d=1$}

\addplot[blue!60!black,dashdotted,thick]
    {gapH(x,2)/\Ztwo};
\addlegendentry{$d=2$}

\addplot[blue!80!black,solid,very thick]
    {gapH(x,4)/\Zfour};
\addlegendentry{$d=4$}

\end{groupplot}
\end{tikzpicture}

\caption{Large-gap concentration and identification
under quartic-exponential errors.}
\label{fig:kotlarski-concentration}

\vspace{0.4em}
\begin{minipage}{0.92\textwidth}
\footnotesize
\emph{Note.}
The idiosyncratic errors have density
$f(\eta)\propto\exp(-\eta^4/4)$.
Panel (a) plots the conditional density of
$\eta_2-d/2$ given $D=d$ for different gaps.
Panel (b) plots the corresponding conditional
density of $X_2-d/2$, taking $\xi\sim N(0,1)$
for illustration. The black dotted curve
represents $f_\xi$. 
\end{minipage}

\end{figure}
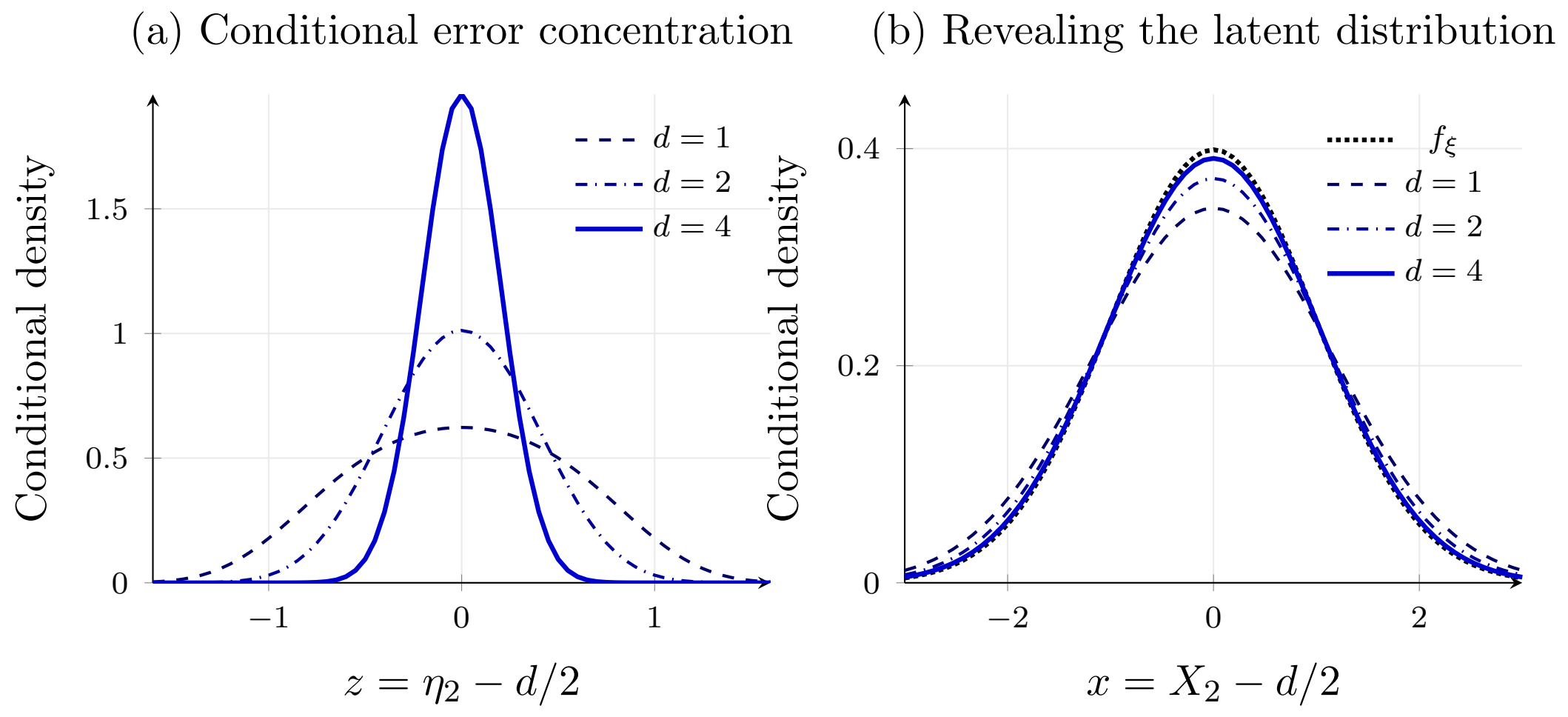

Figure~\ref{fig:kotlarski-concentration} further illustrates
how large-gap conditioning leads to identification.
Panel (a) plots the conditional density of
$\eta_2-d/2$ given $D=d$ for different values of $d$.
As the gap increases, the distribution becomes
increasingly concentrated around zero.
Panel (b) shows the corresponding conditional density
of $X_2-d/2$, taking $\xi\sim N(0,1)$ for illustration.
As the idiosyncratic component concentrates, the
conditional distribution of the recentered measurement
approaches that of $\xi$. Together with
Figure~\ref{fig:kotlarski-large-gap-contour},
this illustrates the key identification mechanism:
large-gap conditioning makes the idiosyncratic component
asymptotically deterministic, allowing the observed
measurement to reveal the latent distribution up to location.

\FloatBarrier
\subsection{General Idea with Order Statistics} 

More generally, our identification argument exploits the spacing between the two observed
order statistics. Define
\[
D=X_{(s)}-X_{(r)}
  =\eta_{(s)}-\eta_{(r)}.
\]
The common component \(\xi\) therefore disappears from the gap. 
Importantly, our identification argument exploits the joint
distribution of $(D,X_{(s)})$, equivalently that of
$(X_{(r)},X_{(s)})$, rather than the marginal distribution
of $D$ alone. In particular, we use the conditional
distribution of $X_{(s)}$ given $D=d$ as the gap becomes
large. This additional information distinguishes our
approach from identification based solely on the spacing
distribution, which is generally insufficient, as
illustrated by \citet{rossberg1972characterization}. The key idea is to study the conditional distribution of an idiosyncratic order statistic when this gap becomes extreme. 

Let
\[
Y_d\sim \eta_{(s)}\mid D=d.
\]
For expositional simplicity, we focus on adjacent order statistics, i.e., $s = r+1$, in the main analysis. Section \ref{sec:extend} discusses how the argument can be extended to nonconsecutive order statistics. In the adjacent case, the conditional density of \(Y_d\) is
proportional to
\[
K_d(y)
=
F(y-d)^{r-1}
\{1-F(y)\}^{n-r-1}
f(y-d)f(y),
\]
(see, e.g., \citet{arnold2008first} and \citet{LuoXiao2023}). The factors $f(y-d)f(y)$ describe the densities
at the two observed order statistics, while
$F(y-d)^{r-1}$ and $\{1-F(y)\}^{n-r-1}$
account for the unobserved observations lying
below and above them. The key distinction from
the classical two-measurement model is the
presence of these additional ranking factors.

Define
\[
Q_d(y)=-\log K_d(y).
\]
The large-gap argument asks what values of \(y\) can plausibly generate
\(D=d\) when \(d\) is extreme. Under suitable support and tail conditions,
the conditional distribution of \(Y_d\) becomes localized around a
deterministic location \(y_d\). In the thinner-than-Gaussian full-support
case studied below,
\[
Y_d-y_d\xrightarrow{p}0,
\qquad d\to\infty.
\]
That is, large-gap conditioning localizes the relevant idiosyncratic order statistic around a deterministic candidate-specific location $y_d$.

The centering location $y_d$ generally depends on
the unknown idiosyncratic distribution and is therefore
not directly observable. Nevertheless, as shown below,
observationally equivalent models can have different
large-gap locations only up to an asymptotically
constant shift. 
Consequently, concentration identifies the latent distribution
up to location without requiring $y_d$ to be known.
To see this, note that conditional on $D=d$,
\[
X_{(s)}\overset{d}{=}\xi+Y_d,
\]
where $\xi$ and $Y_d$ are independent.
Since $Y_d-y_d\xrightarrow{p}0$ as $d\to\infty$,
the conditional distribution of $X_{(s)}-y_d$
converges to that of $\xi$. Thus, large-gap
conditioning separates the idiosyncratic component
from the remaining variation in the observed order
statistic, revealing the common latent distribution
up to location.

The same logic applies across different support configurations. With compact
support, an extreme feasible gap pushes the two idiosyncratic order statistics
toward opposite support boundaries. With one-sided bounded support, one order
statistic is localized at the finite boundary while the other moves into the
unbounded tail. With full support, both locations are determined by tail
geometry. In the next section, we first develop the full-support argument and then show how the
same large-gap logic specializes to the other cases.

\section{Main Results}  \label{sec: main}

This section formalizes the large-gap identification
argument for distributions with two unbounded tails.
We first establish point concentration under a
sufficient tail-curvature condition and then show
that this concentration yields nonparametric
identification of the common and idiosyncratic
distributions. We conclude with extensions to
nonconsecutive order statistics and one-sided
bounded support.

\subsection{Identification under Thinner-than-Gaussian Tails}

Write the right and left tail potentials as
\[
f(z)=\exp\{-U_+(z)\},
\qquad
f(-z)=\exp\{-U_-(z)\},
\qquad z>L_0.
\]
For \(L\ge L_0\) and \(d>2L\), define the large-gap interval
\[
I_{L,d}=[L,d-L].
\]
If \(y\in I_{L,d}\), then the upper order statistic \(y\) is in the right tail,
and the lower order statistic \(y-d=-(d-y)\) is in the left tail.

\begin{assumption}[Regular thinner-than-Gaussian tails] \label{tail}
There exists \(L_0<\infty\) such that \(U_+\) and \(U_-\) are \(C^2\),
strictly increasing, and convex on \([L_0,\infty)\). Moreover,
\[
U_+''(z)\to\infty,
\qquad
U_-''(z)\to\infty,
\qquad z\to\infty.
\]
\end{assumption}

The term ``thinner-than-Gaussian'' reflects the fact that
the negative log-density of a Gaussian distribution is
quadratic, with constant second derivative. Our assumption
requires this curvature to diverge in both tails, implying
that the density decays faster than any Gaussian density.

Assumption~\ref{tail} restricts only the tail behavior,
allowing the density to have an arbitrary shape over
its central region. For example, it is satisfied by
densities with tails proportional to
\[
\exp\{-|z|^\alpha\},
\qquad \alpha>2.
\]
The condition is sufficient rather than necessary for
concentration. Its intuition is simple: as both tails
become increasingly curved, deviations from the most
likely configuration of the idiosyncratic order statistics
become increasingly costly under a large gap.

To formalize this intuition, we first establish that
the conditional potential $Q_d$ has a unique interior
minimizer on the large-gap interval $I_{L,d}$.
Importantly, this minimizer does not depend on the
choice of cutoff $L$.

\begin{lemma}[Interior mode] \label{mode}
There exists \(\bar L\ge L_0\) such that, for every fixed
\(L\ge \bar L\), and all sufficiently large \(d\), \(Q_d\) has a unique
minimizer on \(I_{L,d}\). In fact, this minimizer does not depend on the fixed
cutoff \(L\). We denote the common minimizer by \(y_d\). It satisfies
\[
y_d\to\infty,
\qquad
d-y_d\to\infty.
\]
\end{lemma}

\begin{proof}[Proof sketch]

The argument relies on Assumption~\ref{tail}. By
Lemma~\ref{logconcavity}, for sufficiently large $L$,
\[
Q_d''(y)\ge
U_-''(d-y)+U_+''(y)>0,
\qquad y\in I_{L,d}.
\]
Thus $Q_d$ is strictly convex on $I_{L,d}$
and has at most one minimizer.

Moreover, for any fixed sufficiently large $L$,
\[
Q_d'(L)\to-\infty,
\qquad
Q_d'(d-L)\to+\infty
\]
as $d\to\infty$. Hence, for sufficiently large $d$,
$Q_d'$ has a unique zero in $I_{L,d}$,
which is the unique interior minimizer.

Finally, we show that this minimizer is independent
of $L$. Consider any two fixed cutoffs
$L_2>L_1\ge\bar L$. Since
$I_{L_2,d}\subset I_{L_1,d}$, the interior
minimizers on both intervals are zeros of the same
function $Q_d'$. Strict convexity on the larger
interval implies that they must coincide.
Since $L$ can be chosen arbitrarily large,
both $y_d$ and $d-y_d$ diverge as $d\to\infty$.
The details are provided in
Appendix~\ref{app:mode-proof}.
\end{proof}

Lemma~\ref{mode} establishes a well-defined, gap-dependent
location $y_d$ around which the conditional distribution
may concentrate. To establish point concentration, we need
two additional properties. First, the curvature of $Q_d$
on $I_{L,d}$ becomes arbitrarily large as $L$ increases,
making deviations from $y_d$ increasingly unlikely.
Second, the conditional probability outside $I_{L,d}$
vanishes as $d\to\infty$ for any fixed sufficiently
large $L$. Together, these properties imply that the
conditional distribution collapses to $y_d$, as
formalized in the following theorem.

\begin{theorem}[Point Concentration] \label{main}
Under Assumption \ref{tail},
\[
Y_d-y_d\pto0.
\]
\end{theorem}

\begin{proof}[Proof sketch]
The argument combines increasing curvature with
exterior localization. By Lemma~\ref{logconcavity},
the conditional potential satisfies
\[
Q_d''(y)\ge \lambda_L,
\qquad y\in I_{L,d},
\]
where $\lambda_L\to\infty$ as $L\to\infty$.
Using the interior minimizer $y_d$ from
Lemma~\ref{mode}, convexity yields, for any
$\varepsilon>0$,
\[
\Pr(|Y_d-y_d|>\varepsilon)
\le
\Pr(Y_d\notin I_{L,d})
+\frac{2}{\lambda_L\varepsilon^2}.
\]
By Lemma~\ref{local}, the first term vanishes
as $d\to\infty$ for fixed $L$.
Letting $L\to\infty$ then establishes
point concentration. The full proof is
provided in Appendix~\ref{app:concentration}.
\end{proof}

The concentration result provides the key to identification.
Conditional on an increasingly large gap, the idiosyncratic
order statistic becomes asymptotically deterministic, so
the remaining variation in the corresponding observed order
statistic reveals the distribution of the common latent
component up to location. Once this distribution is identified,
the marginal distribution of an observed order statistic
allows us to recover its idiosyncratic component through
deconvolution. The parent idiosyncratic distribution then
follows from the known order-statistic relationship, and
the location normalization resolves the remaining ambiguity.
The following theorem formalizes these two steps.

\begin{theorem}
\label{thm:fullid}
Under Assumption \ref{tail},  the distributions of \(\xi\) and \(\eta_j\) are identified from the joint distribution of $(X_{(r)},X_{(r+1)}).$
\end{theorem}

The formal argument is given in Appendix~\ref{app:identification}. An important feature of Theorem~\ref{thm:fullid} is that
identification does not require the characteristic function
of the common latent component to be everywhere nonzero.
It suffices to perform deconvolution in a neighborhood of
zero, where this characteristic function is necessarily
nonzero. Under Assumption~\ref{tail}, the characteristic
function of the idiosyncratic order statistic admits an
entire analytic extension, so identification on this
neighborhood extends to the entire real line.
Having established full identification under the
thinner-than-Gaussian tail condition, we next consider
extensions to other pairs of order statistics and
support configurations.

\subsection{Extensions} \label{sec:extend}

The large-gap approach extends to nonconsecutive order
statistics and to distributions with a finite support boundary.
The underlying identification principle remains the same,
although the mechanism generating concentration differs.
We briefly discuss these extensions below.

\paragraph{Nonconsecutive Order Statistics.}

The main results extend to any pair of order statistics,
not necessarily consecutive. For $s>r+1$, the conditional
kernel of $\eta_{(s)}$ given
$D=\eta_{(s)}-\eta_{(r)}=d$ contains the additional factor
\[
\{F(y)-F(y-d)\}^{s-r-1}.
\]
On the large-gap interval $I_{L,d}$, this factor is
bounded away from zero, since
\[
F(y)-F(y-d)
\ge F(L)-F(-L)>0.
\]
Consequently, the additional factor does not prevent
concentration. This yields the following extension of
our main identification result.

\begin{corollary}
\label{cor:nonconsecutive}
Under Assumption~\ref{tail}, for any $1\le r<s\le n$, the joint distribution of $(X_{(r)},X_{(s)})$ identifies the distributions of both $\xi$ and $\eta_j$. 
\end{corollary}

The proof is given in Appendix~\ref{app:nonconsecutive}.
The key observation is that the conditional kernel can
be compared with that of two adjacent order statistics
from a smaller sample. The additional middle factor
is uniformly bounded away from zero on the large-gap
interval, allowing the point-concentration argument
to carry over.

\paragraph{One-Sided Bounded Support.}

The large-gap identification principle also applies when
the idiosyncratic distribution has a finite support boundary,
as in \citet{ChoLuoXiao2024}. Suppose, for example, that
$\eta_j$ has support $[\underline{\eta},\infty)$.
Conditioning on a large gap pushes the lower order statistic
toward the finite boundary, while the upper order statistic
moves into the right tail. Unlike the full-support case,
this localization can occur under a weaker tail condition
than Assumption~\ref{tail}, as shown in the following result.

\begin{corollary}
\label{prop:one-sided}
Suppose the idiosyncratic distribution has support
$[\underline{\eta},\infty)$ and, in the right tail,
$f(z)=\exp\{-U_+(z)\}$ with
\[
U_+'(z)\to\infty
\qquad\text{as }z\to\infty.
\]
For adjacent order statistics, let
$D=X_{(r+1)}-X_{(r)}$. Then
\[
\eta_{(r)}\mid D=d
\xrightarrow{p}\underline{\eta},
\qquad d\to\infty.
\]
Consequently,
\[
X_{(r)}-\underline{\eta}\mid D=d
\Rightarrow \xi.
\]
\end{corollary}

The proof is provided in Appendix~\ref{app:one-sided}.
This result highlights the distinction between the two
localization mechanisms. With a finite lower boundary,
the lower order statistic concentrates at a fixed location,
and the upper order statistic absorbs nearly the entire
additional gap. With two unbounded tails, the large gap
instead determines an endogenous location through tail
curvature. In either case, concentration reveals the
distribution of the common latent component up to location.
Moreover, the finite endpoint need not be known a priori:
observationally equivalent candidate endpoints can differ
only by a location shift, which is resolved by the location
normalization on the idiosyncratic errors.

\section{Conclusion}

This paper develops a large-gap approach to nonparametric deconvolution that applies to both classical repeated-measurement models and settings where only order statistics are observed. The key insight is that within-unit gaps eliminate the common latent component, while conditioning on an extreme gap restricts the locations of the idiosyncratic components through support or tail geometry. After appropriate recentering, the observed measurement reveals the latent distribution up to location, providing the basis for recovering the idiosyncratic distribution through deconvolution. This approach provides a unified geometric explanation for identification from repeated measurements and order statistics, and establishes new identification results for order-statistic models with unbounded support.

We establish point concentration under regular thinner-than-Gaussian tails. More generally, the large-gap perspective suggests that tail geometry may govern not only whether concentration occurs, but also the rate of concentration and the limiting shape of the conditional distribution around its large-gap location. While our main results focus on the increasing-curvature regime, where the conditional distribution collapses to a point, extending the analysis to other tail regimes may yield a broader characterization of conditional order statistics under extreme spacings and new identification results. Finally, although large gaps provide the key to identification, empirical estimation need not rely exclusively on extreme observations and can instead exploit the full joint distribution of the observed measurements or order statistics.

\newpage

\bibliography{ref_gap}

\newpage

\appendix

\setcounter{lemma}{0}
\renewcommand{\thelemma}{\Alph{section}\arabic{lemma}}
\counterwithin{lemma}{section}

\section{Proofs}

\subsection{Proof of Lemma~\ref{mode}}
\label{app:mode-proof}

\begin{proof}
Choose \(\bar L\ge L_0\) large enough that \(\lambda_{\bar L} >0.\) Then, for every \(L\ge \bar L\), Lemma \ref{logconcavity} implies that \(Q_d\) is strictly convex on $I_{L,d}$. Thus \(Q_d\) has at most one minimizer on \(I_{L,d}\).

First, we show that such a minimizer exists and is interior. For \(y\in I_{L,d}\),
differentiating the conditional potential $Q_d$ gives
\[
\begin{aligned}
Q_d'(y)
=&
-(r-1)\frac{f(y-d)}{F(y-d)}
+(n-r-1)\frac{f(y)}{1-F(y)} 
-U_-'(d-y)+U_+'(y).
\end{aligned}
\]
At the left endpoint \(y=L\), the terms involving \(L\)
itself remain fixed as \(d\to\infty\), while
\[
-(r-1)\frac{f(y-d)}{F(y-d)} < 0, \qquad
U_-'(d-L)\to\infty.
\]
Hence
\[
Q_d'(L)\to-\infty.
\]
At the right endpoint \(y=d-L\), the terms involving the left-tail cutoff \(L\)
remain fixed, while
\[
U_+'(d-L)\to\infty.
\]
Hence
\[
Q_d'(d-L)\to+\infty.
\]
Therefore, for all sufficiently large \(d\),
\[
Q_d'(L)<0,
\qquad
Q_d'(d-L)>0.
\]
By continuity, \(Q_d'\) has a zero in \((L,d-L)\). By strict convexity on
\(I_{L,d}\), this zero is unique and is the unique minimizer of \(Q_d\) on
\(I_{L,d}\).

Second, we show that the minimizer does not depend on the fixed cutoff \(L\). Take two fixed cutoffs \(L_2>L_1\ge \bar L\). Take \(d\) sufficiently large so that \(d>2L_2\) and the preceding existence argument applies to both cutoffs. Then \(Q_d\) has a unique interior minimizer on each of the intervals \(I_{L_1,d}\) and \(I_{L_2,d}\). Since $I_{L_2,d}\subset I_{L_1,d}$, the two minimizers must coincide. Hence the minimizer is independent of the particular fixed cutoff \(L\), once \(d\) is large.

Finally, because $L<y_d<d-L$ and \(L\) is arbitrary, this implies
\[
y_d\to\infty,
\qquad
d-y_d\to\infty.
\]
\end{proof}

\subsection{Proof of Theorem \ref{main}}  \label{app:concentration}

\begin{proof}
Fix \(\eps>0\). Choose \(L\ge \bar L\) large. By Lemma \ref{mode} applied with cutoff \(L+\eps\), for all sufficiently
large \(d\), $L+\eps<y_d<d-L-\eps.$ Therefore
\[
[y_d-\eps,y_d+\eps]\subset I_{L,d}.
\]

We first bound the left side of the conditional distribution inside
\(I_{L,d}\). On \(I_{L,d}\), Lemma \ref{logconcavity} gives
\[
Q_d''(y)\ge \lambda_L.
\] 
Since \(Q_d'(y_d)=0\),
\[
Q_d'(y_d-\eps)
=
-\int_{y_d-\eps}^{y_d}Q_d''(u)\,du
\le
-\lambda_L\eps.
\]
By convexity on \(I_{L,d}\), for \(y\in[L,y_d-\eps]\),
\[
Q_d(y)
\ge
Q_d(y_d-\eps)
+
Q_d'(y_d-\eps)\{y-(y_d-\eps)\}.
\]
Therefore,
\[
\int_L^{y_d-\eps} e^{-Q_d(y)}\,dy
\le
\frac{e^{-Q_d(y_d-\eps)}}{\lambda_L\eps}.
\]

On the other hand, \(Q_d\) is decreasing on \([y_d-\eps,y_d]\). Hence
\[
\int_{y_d-\eps}^{y_d} e^{-Q_d(y)}\,dy
\ge
\eps e^{-Q_d(y_d-\eps)}.
\]
Since
\[
\Pr\{Y_d\in I_{L,d},\,Y_d\le y_d-\eps\}
=
\frac{
\int_L^{y_d-\eps} e^{-Q_d(y)}\,dy
}{
\int K_d(u)\,du
},
\]
and since the denominator satisfies
\[
\int K_d(u)\,du
\ge
\int_{y_d-\eps}^{y_d} e^{-Q_d(y)}\,dy,
\]
we obtain
\[
\Pr\{Y_d\in I_{L,d},\,Y_d\le y_d-\eps\}
\le
\frac{
e^{-Q_d(y_d-\eps)}/(\lambda_L\eps)
}{
\eps e^{-Q_d(y_d-\eps)}
}
=
\frac{1}{\lambda_L\eps^2}.
\]

The same argument on the right gives
\[
\Pr\{Y_d\in I_{L,d},\,Y_d\ge y_d+\eps\}
\le
\frac{1}{\lambda_L\eps^2}.
\]
Therefore,
\[
\Pr\{|Y_d-y_d|>\eps\}
\le
\Pr\{Y_d\notin I_{L,d}\}
+
\frac{2}{\lambda_L\eps^2}.
\]
For fixed \(L\), Lemma \ref{local} implies
\[
\Pr\{Y_d\notin I_{L,d}\}\to0
\qquad(d\to\infty).
\]
Thus
\[
\limsup_{d\to\infty}
\Pr\{|Y_d-y_d|>\eps\}
\le
\frac{2}{\lambda_L\eps^2}.
\]
Finally let \(L\to\infty\). Since \(\lambda_L\to\infty\), the right-hand side
vanishes. Hence
\[
Y_d-y_d\pto0.
\]
\end{proof}

\subsection{Proof of Theorem \ref{thm:fullid}}  \label{app:identification}

\begin{proof}
Suppose that two candidate models
\[
(F,F_{\xi})
\qquad\text{and}\qquad
(G,G_{\xi})
\]
are observationally equivalent. By Lemma \ref{IDxi}, there
exists a constant \(c\) such that
\[
\xi_G\stackrel{d}{=}\xi_F-c.
\]

Since
\[
X_{(r+1)}=\xi+\eta_{(r+1)}
\]
and \(\xi\) is independent of the idiosyncratic components, observational
equivalence implies
\[
\phi_{\xi_F}(t)\phi_{\eta_{(r+1),F}}(t)
=
\phi_{\xi_G}(t)\phi_{\eta_{(r+1),G}}(t).
\]
Because
\[
\xi_G\stackrel{d}{=}\xi_F-c,
\]
we have
\[
\phi_{\xi_G}(t)
=
e^{-itc}\phi_{\xi_F}(t).
\]
Hence
\[
\phi_{\xi_F}(t)\phi_{\eta_{(r+1),F}}(t)
=
e^{-itc}\phi_{\xi_F}(t)
\phi_{\eta_{(r+1),G}}(t).
\]

Since \(\phi_{\xi_F}(0)=1\) and characteristic functions are continuous,
there exists \(\delta>0\) such that
\[
\phi_{\xi_F}(t)\neq0,
\qquad |t|<\delta.
\]
Therefore,
\[
\phi_{\eta_{(r+1),G}}(t)
=
e^{itc}\phi_{\eta_{(r+1),F}}(t),
\qquad |t|<\delta.
\]

Assumption \ref{tail} implies that the idiosyncratic error has exponential
moments of every order. Indeed, since
\[
U_+''(z)\to\infty,
\qquad
U_-''(z)\to\infty,
\]
we also have
\[
U_+'(z)\to\infty,
\qquad
U_-'(z)\to\infty.
\]
Thus, for every \(a>0\), both tails of \(f\) are eventually bounded by a
multiple of \(e^{-a|z|}\), so that
\[
\E[e^{a|\eta_j|}]<\infty.
\]
Since
\[
|\eta_{(r+1)}|
\leq
\max_{1\leq j\leq n}|\eta_j|,
\]
it follows that
\[
\E[e^{a|\eta_{(r+1)}|}]<\infty
\qquad\text{for every }a>0.
\]
Hence the characteristic function of \(\eta_{(r+1)}\) admits an entire
extension to the complex plane.

The two entire functions
\[
\phi_{\eta_{(r+1),G}}(z)
\qquad\text{and}\qquad
e^{icz}\phi_{\eta_{(r+1),F}}(z)
\]
therefore agree on an interval containing zero. By the identity theorem,
they agree everywhere, and hence
\[
\eta_{(r+1),G}
\stackrel{d}{=}
\eta_{(r+1),F}+c.
\]

The distribution of an order statistic is a one-to-one transformation of
the parent distribution; see \citet{AtheyHaile2002}. Therefore,
\[
G(x)=F(x-c),
\]
so the parent idiosyncratic distribution is identified up to the same
location shift \(c\).

Finally, both candidate distributions satisfy
\[
\E[\eta_j]=0.
\]
Therefore,
\[
0
=
\E_G[\eta_j]
=
\E_F[\eta_j]+c
=
c.
\]
Hence \(c=0\), so
\[
G=F
\]
and
\[
F_{\xi_G}=F_{\xi_F}.
\]
Thus the distributions of both \(\xi\) and \(\eta_j\) are identified.
\end{proof}

\subsection{Supplementary Lemmas}

\begin{lemma}[Log-concavity and curvature on the large-gap interval] \label{logconcavity}
For fixed \(L\ge L_0\) and all \(d>2L\), \(K_d(\cdot)\) is log-concave on
\(I_{L,d}\). 
Then, for \(y\in I_{L,d}\),
\[
Q_d''(y)
\ge
U_-''(d-y)+U_+''(y).
\]
\end{lemma}

\begin{proof}
On \(I_{L,d}\), write
\[
y-d=-(d-y).
\]
The density terms in \(\log K_d\) are
\[
\log f(y-d)+\log f(y)
=
-U_-(d-y)-U_+(y).
\]
Their contribution to \(Q_d''\) is exactly
\[
U_-''(d-y)+U_+''(y).
\]

The remaining terms are tail-probability terms:
\[
(r-1)\log F(y-d)
\quad\text{and}\quad
(n-r-1)\log\{1-F(y)\}.
\]
A standard one-dimensional implication of log-concavity is that the integral of a log-concave tail density is log-concave. Thus, both
\[
F(y-d)=\int_{d-y}^{\infty}e^{-U_-(u)}\,du
\]
and
\[
1-F(y)=\int_y^{\infty}e^{-U_+(u)}\,du
\]
are log-concave functions of \(y\) on \(I_{L,d}\). Hence the tail-probability terms add concavity to \(\log K_d\) 
\[
Q_d''(y)
\ge
U_-''(d-y)+U_+''(y).
\]
Note that \(\lambda_L
=
\inf_{z\ge L}U_-''(z)+\inf_{z\ge L}U_+''(z) \to\infty\) follows from
Assumption \ref{tail}. 
\end{proof}

\begin{lemma}[Exterior localization] \label{local}
Fix \(L\ge L_0\), then
\[
\Pr(Y_d\le L)\to 0
\qquad\text{and}\qquad
\Pr(Y_d\ge d-L)\to 0
\]
as \(d\to\infty\). Hence
\[
\Pr(Y_d\notin I_{L,d})\to0.
\]
\end{lemma}

\begin{proof}
We only prove the lower-tail case.\footnote{The upper-tail case is symmetric comparing the exterior region \(y\ge d-L\) with the interior strip \([d-L-2,d-L-1]\).} The argument compares the unnormalized conditional mass in the exterior region
\((-\infty,L]\) with the unnormalized mass on the fixed interior strip
\([L+1,L+2]\). The lower endpoint is \(y-d=-(d-y)\), so the relevant left-tail
distance is \(d-y\). If \(y\le L\), then \(d-y\ge d-L\). By contrast, if
\(y\in[L+1,L+2]\), then \(d-y\le d-L-1\). Thus the fixed interior strip reduces
the required left-tail distance by at least one unit. Since the left-tail slope
\(U_-'(z)\) diverges, this one-unit reduction lowers the tail cost by an amount
that tends to infinity. As a result, $\int_{-\infty}^{L}K_d(y)\,dy
$ is negligible relative to $\int_{L+1}^{L+2}K_d(y)\,dy
$ as \(d\to\infty\).

First consider the exterior region \(y\le L\). Then \(d-y\ge d-L\). For all sufficiently large \(d\), these arguments are in the left-tail region.
Since \(U_-\) is increasing,
\[
f(y-d)
=
f(-(d-y))
=
\exp\{-U_-(d-y)\}
\le
\exp\{-U_-(d-L)\}.
\]
Also,
\[
F(y-d)
=
F(-(d-y))
\le
F(-(d-L)).
\]
Using the conditional kernel \(K_d(y)
=
F(y-d)^{r-1}
\{1-F(y)\}^{n-r-1}
f(y-d)f(y)\), \(\{1-F(y)\}^{n-r-1}\le 1\) and \(\int_{-\infty}^{L}f(y)\,dy\le 1\), we obtain
\[
\int_{-\infty}^{L}K_d(y)\,dy
\le
F(-(d-L))^{r-1}
\exp\{-U_-(d-L)\}.
\]

Second consider the interior strip $[L+1,L+2],$ where the terms involving \(y\) itself are bounded away from zero. Hence there exists \(c_L>0\), depending only on \(L\), such that
\[
\{1-F(y)\}^{n-r-1}f(y)\ge c_L.
\]
Moreover,
\[
F(y-d)
=
F(-(d-y))
\ge
F(-(d-L-1)),
\]
and
\[
f(y-d)
=
\exp\{-U_-(d-y)\}
\ge
\exp\{-U_-(d-L-1)\}.
\]
It follows that
\[
\int K_d(u)\,du
\ge
\int_{L+1}^{L+2}K_d(y)\,dy
\ge
c_L
F(-(d-L-1))^{r-1}
\exp\{-U_-(d-L-1)\}.
\]

Combining the upper bound on the numerator and the lower bound on the
denominator gives
\[
\Pr(Y_d\le L) =
\frac{\int_{-\infty}^{L}K_d(y)\,dy}
{\int K_d(u)\,du}
\le
c_L^{-1}
\left\{
\frac{F(-(d-L))}{F(-(d-L-1))}
\right\}^{r-1}
\exp\left\{-[U_-(d-L)-U_-(d-L-1)]\right\}.
\]
Since
\[
F(-(d-L))\le F(-(d-L-1)),
\]
we have
\[
\Pr(Y_d\le L)
\le
c_L^{-1}
\exp\left\{-[U_-(d-L)-U_-(d-L-1)]\right\}.
\]

Finally,
\[
U_-(d-L)-U_-(d-L-1)
=
\int_{d-L-1}^{d-L}U_-'(u)\,du.
\]
Because \(U_-''(z)\to\infty\), we have \(U_-'(z)\to\infty\). Hence
\[
U_-(d-L)-U_-(d-L-1)\to\infty.
\]
Therefore
\[
\Pr(Y_d\le L)\to0.
\]
\end{proof}

\begin{lemma} \label{IDxi}
Under Assumption \ref{tail},  the joint distribution
of $(X_{(r)},X_{(r+1)})$ identifies the distribution
of $\xi$ up to location.
\end{lemma}

\begin{proof}
Suppose $(F,F_{\xi})$ and $(G,G_{\xi})$ are observationally equivalent, and
let $y_F(d)$ and $y_G(d)$ denote their respective large-gap locations. Since
the conditional distribution of $X_{(s)}$ given $D=d$ is the same under the
two models,
\[
X_{(s)}-y_F(d)\mid D=d
\Rightarrow \xi_F,
\]
and
\[
X_{(s)}-y_G(d)\mid D=d
\Rightarrow \xi_G.
\]

We first show that
\[
y_G(d)-y_F(d)
\]
must be bounded. Since
\[
X_{(s)}-y_F(d)\mid D=d \Rightarrow \xi_F
\]
and \(\xi_F\) has a proper probability distribution, for any
\(\varepsilon>0\) there exists a finite \(M\) such that, for all sufficiently
large \(d\),
\[
\Pr\left(
\left|X_{(s)}-y_F(d)\right|\leq M
\,\middle|\, D=d
\right)
>1-\varepsilon.
\]
Similarly, because
\[
X_{(s)}-y_G(d)\mid D=d \Rightarrow \xi_G,
\]
we can choose \(M\) sufficiently large so that, for all sufficiently large
\(d\),
\[
\Pr\left(
\left|X_{(s)}-y_G(d)\right|\leq M
\,\middle|\, D=d
\right)
>1-\varepsilon.
\]

Suppose, to the contrary, that \(y_G(d)-y_F(d)\) is unbounded. Then there
exists a subsequence \(d_k\to\infty\) such that
\[
\left|y_G(d_k)-y_F(d_k)\right|>2M.
\]
But
\[
X_{(s)}-y_G(d_k)
=
\left\{X_{(s)}-y_F(d_k)\right\}
-
\left\{y_G(d_k)-y_F(d_k)\right\}.
\]
Hence, whenever
\[
\left|X_{(s)}-y_F(d_k)\right|\leq M,
\]
the triangle inequality implies
\[
\begin{aligned}
\left|X_{(s)}-y_G(d_k)\right|
&\geq
\left|y_G(d_k)-y_F(d_k)\right|
-
\left|X_{(s)}-y_F(d_k)\right|  \\
&> 2M-M=M.
\end{aligned}
\]
Therefore,
\[
\Pr\left(
\left|X_{(s)}-y_G(d_k)\right|>M
\,\middle|\,D=d_k
\right)
>
1-\varepsilon.
\]
Choosing \(\varepsilon<1/2\), this contradicts the fact that, for sufficiently
large \(d_k\),
\[
\Pr\left(
\left|X_{(s)}-y_G(d_k)\right|\leq M
\,\middle|\,D=d_k
\right)
>1-\varepsilon.
\]
Thus \(y_G(d)-y_F(d)\) must be bounded.

Now consider any subsequence such that
\[
y_G(d_k)-y_F(d_k)\longrightarrow c.
\]
Along this subsequence,
\[
X_{(s)}-y_G(d_k)
=
\{X_{(s)}-y_F(d_k)\}
-
\{y_G(d_k)-y_F(d_k)\}.
\]
Therefore, by Slutsky's theorem,
\[
X_{(s)}-y_G(d_k)\mid D=d_k
\Rightarrow
\xi_F-c.
\]
But the left-hand side also converges to \(\xi_G\). Hence
\[
\xi_G\stackrel{d}{=}\xi_F-c.
\]

It remains to show that the subsequential limit \(c\) is unique. Suppose that
another subsequence satisfies
\[
y_G(d_\ell)-y_F(d_\ell)\longrightarrow c'.
\]
Repeating the same argument gives
\[
\xi_G\stackrel{d}{=}\xi_F-c'.
\]
Therefore,
\[
\xi_F-c\stackrel{d}{=}\xi_F-c'.
\]
If \(c\neq c'\), this would imply
\[
\xi_F\stackrel{d}{=}\xi_F+(c-c'),
\]
so the distribution of \(\xi_F\) would be invariant to a nonzero location
shift. This is impossible for a proper probability distribution. Hence
\(c=c'\).

Since \(y_G(d)-y_F(d)\) is bounded and every convergent subsequence has the
same limit, it follows that
\[
y_G(d)-y_F(d)\longrightarrow c.
\]
Consequently,
\[
\xi_G\stackrel{d}{=}\xi_F-c.
\]
Thus the distribution of the common latent component is identified up to
location.
\end{proof}

\subsection{Nonconsecutive Order Statistics}
\label{app:nonconsecutive}

This subsection shows that the large-gap concentration argument extends to nonconsecutive order statistics. Let \(1\le r<s\le n\), with \(s>r+1\), and define
$$
D=\eta_{(s)}-\eta_{(r)},
\qquad
Y_d\sim \eta_{(s)}\mid D=d.
$$

Conditional on \(D=d\), the density of \(Y_d\) is proportional to

$$
K_d^{r,s}(y)
=
F(y-d)^{r-1}
\{F(y)-F(y-d)\}^{s-r-1}
\{1-F(y)\}^{n-s}
f(y-d)f(y).
$$

Let
$$
k=s-r-1
$$
and define the benchmark kernel
$$
\widetilde K_d(y)
=
F(y-d)^{r-1}
\{1-F(y)\}^{n-s}
f(y-d)f(y).
$$

Since
$$
n-s=(n-k)-r-1,
$$
\(\widetilde K_d\) is exactly the conditional kernel for two adjacent order
statistics of ranks \(r\) and \(r+1\) from a sample of size \(n-k\).
Therefore, under Assumption~\ref{tail}, Theorem~\ref{main} applies to this
benchmark problem. Let \(y_d\) denote its deterministic large-gap location.
If \(\widetilde Y_d\) has density proportional to \(\widetilde K_d\), then

$$
\widetilde Y_d-y_d\xrightarrow{p}0.
$$

Moreover, for every sufficiently large fixed \(L\),

$$
\Pr(\widetilde Y_d\in I_{L,d})\to1.
$$

It remains to show that the additional middle factor in \(K_d^{r,s}\) does
not affect this concentration. For \(y\in I_{L,d}=[L,d-L]\),

$$
F(y)-F(y-d)
\geq
F(L)-F(-L)
=:c_L.
$$

For sufficiently large \(L\), \(c_L>0\), and
$$
c_L\to1
\qquad\text{as }L\to\infty.
$$

Since $F(y)-F(y-d)\leq1,$ we have
$$
K_d^{r,s}(y)\leq \widetilde K_d(y)
$$
for all \(y\), while on \(I_{L,d}\),
$$
K_d^{r,s}(y)
\geq
c_L^k\widetilde K_d(y).
$$

Fix \(\varepsilon>0\). Using the upper bound for the numerator and the lower
bound for the denominator,
$$
\begin{aligned}
\Pr\bigl(|Y_d-y_d|>\varepsilon\bigr)
\leq
c_L^{-k}
\frac{
\displaystyle
\int_{|y-y_d|>\varepsilon}\widetilde K_d(y)\,dy
}{
\displaystyle
\int_{I_{L,d}}\widetilde K_d(y)\,dy
} =
c_L^{-k}
\frac{
\Pr\bigl(|\widetilde Y_d-y_d|>\varepsilon\bigr)
}{
\Pr(\widetilde Y_d\in I_{L,d})
}.
\end{aligned}
$$

The numerator on the right converges to zero, while the denominator converges
to one. Hence

$$
\Pr\bigl(|Y_d-y_d|>\varepsilon\bigr)\to0.
$$

Therefore,

$$
Y_d-y_d\xrightarrow{p}0.
$$

Thus the same large-gap concentration result extends to nonconsecutive order statistics. The additional observations lying between \(\eta_{(r)}\) and \(\eta_{(s)}\) contribute the factor $\{F(y)-F(y-d)\}^{s-r-1},$ but on the large-gap region this factor is uniformly bounded away from zero and approaches one as the two endpoints move into opposite tails. The full identification result then follows by the same arguments as in the proof of Theorem~\ref{thm:fullid}, applied to the order statistic $\eta_{(s)}$.

\subsection{The One-Sided Bounded-Support Case}
\label{app:one-sided}

This subsection establishes the localization result used in the one-sided bounded-support case. We prove the result in three steps.

\medskip
\noindent
\textit{Step 1.}
The joint density of two adjacent order statistics implies that, conditional
on \(D=d\), the density of \(\eta_{(r)}\) is proportional to

$$
K_d(x)
=
F(x)^{r-1}f(x)
f(x+d)\{1-F(x+d)\}^{n-r-1},
\qquad x>\underline{\eta}.
$$

Write

$$
A(x)=F(x)^{r-1}f(x)
$$

and

$$
B(z)=f(z)\{1-F(z)\}^{n-r-1}.
$$

Then

$$
K_d(x)=A(x)B(x+d).
$$

Because \(U_+'(z)\to\infty\), \(U_+'(z)>0\) for all sufficiently large
\(z\), so \(f(z)=\exp\{-U_+(z)\}\) is eventually decreasing. Since
\(1-F(z)\) is also decreasing, \(B(z)\) is eventually decreasing.

\medskip
\noindent
\textit{Step 2.}
Fix \(\varepsilon>0\). Since

$$
\eta_{(r)}\geq\underline{\eta}
$$

almost surely, it suffices to show

$$
\Pr\left(
\eta_{(r)}\geq\underline{\eta}+\varepsilon
\mid D=d
\right)\to0.
$$

Using the conditional kernel,

$$
\Pr\left(
\eta_{(r)}\geq\underline{\eta}+\varepsilon
\mid D=d
\right)
=
\frac{
\displaystyle
\int_{\underline{\eta}+\varepsilon}^{\infty}
A(x)B(x+d)\,dx
}{
\displaystyle
\int_{\underline{\eta}}^{\infty}
A(x)B(x+d)\,dx
}.
\label{A.1}
$$

For all sufficiently large \(d\), monotonicity of \(B\) gives

$$
\int_{\underline{\eta}+\varepsilon}^{\infty}
A(x)B(x+d)\,dx
\leq
B(d+\underline{\eta}+\varepsilon)
\int_{\underline{\eta}+\varepsilon}^{\infty}A(x)\,dx.
\label{A.2}
$$

Moreover,

$$
\int_{\underline{\eta}+\varepsilon}^{\infty}A(x)\,dx
=
\frac{
1-F(\underline{\eta}+\varepsilon)^r
}{r}.
$$

For the denominator, restrict attention to the fixed interval

$$
J_\varepsilon
=
\left[
\underline{\eta}+\frac{\varepsilon}{4},
\underline{\eta}+\frac{\varepsilon}{2}
\right].
$$

Since \(f\) is positive on the interior of its support,

$$
\int_{J_\varepsilon}A(x)\,dx
=
\frac{
F(\underline{\eta}+\varepsilon/2)^r
-
F(\underline{\eta}+\varepsilon/4)^r
}{r}
>0.
$$

Again using monotonicity of \(B\),

$$
\begin{aligned}
\int_{\underline{\eta}}^{\infty}A(x)B(x+d)\,dx
&\geq
\int_{J_\varepsilon}A(x)B(x+d)\,dx\\
&\geq
B\left(d+\underline{\eta}+\frac{\varepsilon}{2}\right)
\int_{J_\varepsilon}A(x)\,dx.
\end{aligned}
$$

Combining the above results,
$$
\Pr\left(
\eta_{(r)}\geq\underline{\eta}+\varepsilon
\mid D=d
\right)
\leq
C_\varepsilon
\frac{
B(d+\underline{\eta}+\varepsilon)
}{
B(d+\underline{\eta}+\varepsilon/2)
},
$$

where

$$
C_\varepsilon
=
\frac{
1-F(\underline{\eta}+\varepsilon)^r
}{
F(\underline{\eta}+\varepsilon/2)^r
-
F(\underline{\eta}+\varepsilon/4)^r
}
<\infty.
$$

Thus it remains only to show that the ratio of the two \(B\)-terms converges
to zero.

\medskip
\noindent
\textit{Step 3.}
By the definition of \(B\),

$$
\frac{
B(d+\underline{\eta}+\varepsilon)
}{
B(d+\underline{\eta}+\varepsilon/2)
}
=
\frac{
f(d+\underline{\eta}+\varepsilon)
}{
f(d+\underline{\eta}+\varepsilon/2)
}
\times
\left[
\frac{
1-F(d+\underline{\eta}+\varepsilon)
}{
1-F(d+\underline{\eta}+\varepsilon/2)
}
\right]^{n-r-1}.
$$

The second factor is at most one. Therefore,

$$
\frac{
B(d+\underline{\eta}+\varepsilon)
}{
B(d+\underline{\eta}+\varepsilon/2)
}
\leq
\frac{
f(d+\underline{\eta}+\varepsilon)
}{
f(d+\underline{\eta}+\varepsilon/2)
}.
$$

Using \(f(z)=\exp\{-U_+(z)\}\) in the upper tail,

$$
\begin{aligned}
\frac{
f(d+\underline{\eta}+\varepsilon)
}{
f(d+\underline{\eta}+\varepsilon/2)
}
&=
\exp\left\{
-\left[
U_+(d+\underline{\eta}+\varepsilon)
-
U_+(d+\underline{\eta}+\varepsilon/2)
\right]
\right\}\\
&=
\exp\left\{
-\int_{d+\underline{\eta}+\varepsilon/2}
       ^{d+\underline{\eta}+\varepsilon}
U_+'(u)\,du
\right\}.
\end{aligned}
$$

The interval of integration has fixed length \(\varepsilon/2\), while its
left endpoint tends to infinity. Since \(U_+'(u)\to\infty\),

$$
\int_{d+\underline{\eta}+\varepsilon/2}
       ^{d+\underline{\eta}+\varepsilon}
U_+'(u)\,du
\to\infty.
$$

Hence

$$
\frac{
B(d+\underline{\eta}+\varepsilon)
}{
B(d+\underline{\eta}+\varepsilon/2)
}
\to0.
$$

Therefore, we have $\Pr\left(
\eta_{(r)}\geq\underline{\eta}+\varepsilon
\mid D=d
\right)\to0$ for every $\epsilon > 0$, i.e., $\eta_{(r)}\mid D=d
\xrightarrow{p}
\underline{\eta}.$

\subsection{Details of the Quartic-Exponential Example}
\label{app:quartic}

We provide the derivations for the example in
Section~\ref{sec:example}.

\paragraph{Conditional density and mode.}
The transformation
$(\eta_1,\eta_2)\mapsto(D,\eta_2)$
has Jacobian of absolute value one. Hence,
conditional on $D=d$, the density of $\eta_2$
is proportional to
\[
f(y-d)f(y)
\propto
\exp\left\{-\frac{(y-d)^4+y^4}{4}\right\}.
\]
The corresponding negative log-kernel is
\[
Q_d(y)=\frac{(y-d)^4+y^4}{4},
\]
with derivatives
\[
Q_d'(y)=(y-d)^3+y^3,\qquad
Q_d''(y)=3(y-d)^2+3y^2>0
\]
for $d>0$. Since $Q_d'(d/2)=0$, the unique
conditional mode is $y_d=d/2$, corresponding
to $(-d/2,d/2)$ on the conditioning line.

\paragraph{Concentration.}
Writing $y=z+d/2$, we have
\[
(y-d)^4+y^4
=2z^4+3d^2z^2+\frac{d^4}{8}.
\]
Thus, the conditional density of
$Z_d=\eta_2-d/2$ given $D=d$ is proportional to
\[
\exp\left\{
-\frac{3d^2z^2}{4}-\frac{z^4}{2}
\right\}.
\]
After the change of variables $w=dz$, the
density of $W_d=dZ_d$ is proportional to
\[
\exp\left\{
-\frac{3w^2}{4}-\frac{w^4}{2d^4}
\right\}.
\]
This expression converges pointwise to
$\exp(-3w^2/4)$ and is bounded above by the
same integrable function. By dominated
convergence, the normalizing constant
converges to $\sqrt{4\pi/3}$, yielding
\[
W_d\xrightarrow{d}N\left(0,\frac23\right).
\]
Consequently, $Z_d=O_p(d^{-1})$ and hence
\[
\eta_2-\frac d2\mid D=d\xrightarrow{p}0.
\]
Finally, independence of $\xi$ and the
idiosyncratic errors implies
\[
X_2-\frac d2\mid D=d\Rightarrow\xi.
\]

\end{document}